\documentclass[prb,aps,twocolumn,showpacs]{revtex4-1}
\usepackage{epsfig}
\usepackage{amsmath}
\usepackage{verbatim}

\begin{document}

\title{Quasi-particle scattering and protected nature of topological
states in a parent topological insulator Bi$_2$Se$_3$}

\author{S. R. Park$^1$, W. S. Jung$^1$, Chul Kim$^1$, D. J. Song$^1$,
C. Kim$^{1,*}$, S. Kimura$^2$, K. D. Lee$^3$ and N. Hur$^3$}

\affiliation{$^1$Institute of Physics and Applied Physics, Yonsei
University, Seoul, Korea}

\affiliation{$^2$UVSOR Facility, Institute for Molecular Science
and The Graduate University for Advanced Studies, Okazaki
444-8585, JAPAN}

\affiliation{$^3$Department of Physics, Inha University, Incheon
402-751, Korea}

\date{\today}

\begin{abstract}
We report on angle resolved photoemission spectroscopic studies on
a parent topological insulator (TI), Bi$_2$Se$_3$. The line width
of the spectral function (inverse of the quasi-particle lifetime)
of the topological metallic (TM) states shows an anomalous
behavior. This behavior can be reasonably accounted for by
assuming decay of the quasi-particles predominantly into bulk
electronic states through electron-electron interaction and defect
scattering. Studies on aged surfaces reveal that topological
metallic states are very much unaffected by the potentials created
by adsorbed atoms or molecules on the surface, indicating that
topological states could be indeed protected against weak
perturbations. 
\end{abstract}
\maketitle

TIs are materials with bulk gaps due to spin-orbit coupling. TIs
are classified into `weak TI' and `strong TI' according to Z$_2$
topological invariants.\cite{Fu1,Fu2} Strong TIs have odd number
of TM Dirac cones on the surface which is robust (protected)
against disorder or impurities.

%Essence of properties of topological metals
TM states realized on the surface of a strong TI is important and
could be useful.\cite{Day,Ramirez} The properties of the TM are
also set by the topological nature of the TI. The essential
properties of TMs can be summarized as follows. First, electron
spins in TM are locked into the momenta, forming spin chiral
states\cite{Hsieh}. Such spin chiral states are also well known in
the field of surface science in terms of Rashba effects in surface
states (for example, Sb (111)\cite{Sugawara}, Bi(111)\cite{Koroteev} and Au
(111)\cite{Reinert} surfaces states). Second, back scattering is
suppressed due to the spin chirality,\cite{Roushan} meaning
relatively long quasi-particle life time. Third, metallic bands
are protected against perturbations to the first order due to the
topological nature. This point has yet to be experimentally
observed.

%Experimental results so far, mostly on the existence/spin chirality
Experimental verification of the novel properties of TM is not
only important in the fundamental aspect but also necessary for
use of these materials for future applications. Due to the surface
nature of the TM states, most of the experimental data thus far
came from angle resolved photoemission
(ARPES)\cite{Hsieh,Xia,Hsieh2,Chen} and to a less degree from
scanning tunnelling microscopy (STM).\cite{Roushan} By using
ARPES, it was shown that there exist odd number of bands crossing
the Fermi level in this class of materials.\cite{Xia,Hsieh2}
Moreover, spin resolved photoemission results show that the
electron spins are indeed locked into the momentum and form spin
chiral states.\cite{Hsieh} As for STM studies, a recent study
shows suppression of back scattering, consistent with spin chiral
states.\cite{Roushan}

%Life time, that is, linesshape
Studies mentioned above are about existence and spin chiral states
of TM and experimental verification of whether topological states
are in general protected or not has not been discussed. Protected
topological states should manifest themselves with long
quasi-particle life time. In that regards, ARPES is an important
tool because long quasi-particle life time should result in a
sharp ARPES line-shape. In spite of the intensive recent efforts
on TM electronic structure studies, photoemission line-shape issue
has not been addressed. This point, however, is important because
recent discussions on the use of TIs for spintronic applications
require very low scattering rate to properly convey the spin
information.

To address the quasi-particle lifetime issue, high resolution
ARPES data on various TIs have to be obtained. We begin the effort
with studies on a parent TI Bi$_2$Se$_3$. The goal of this work is
to analyze the high quality ARPES spectral function in terms of
various scattering channels to see the mechanisms behind the
quasi-particle scattering. In that analysis, the bulk states play
an important role in quasi-particle dynamics in the TM band:
Electrons in TM are strongly scattered to the bulk electronic
states by coupling to phonons, impurity or defect created
potentials and other electrons. In addition, we find the TM states
protected against adsorbate induced disorder potentials.

\begin{figure*}
\centering \epsfxsize=17cm \epsfbox{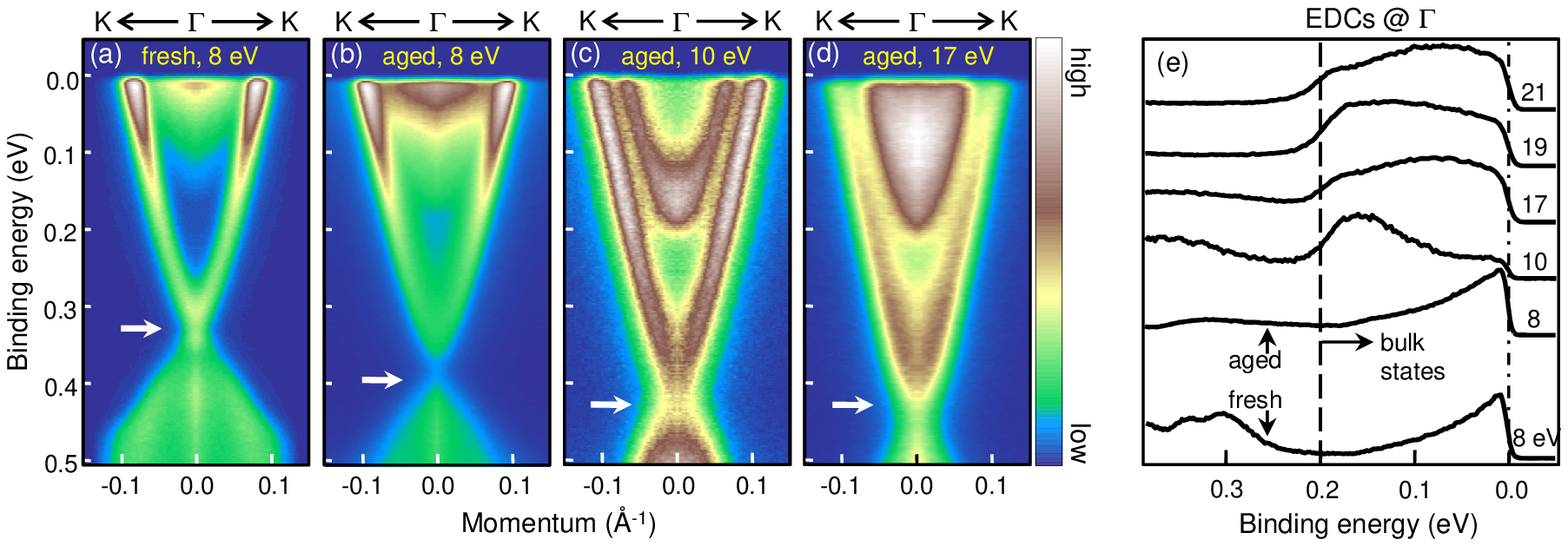} \caption{ARPES data
taken with (a) 8 eV, (b) 8 eV, (b) 10 eV and (c) 17 eV photon
energies along the $\Gamma$ to K points. The data in (a) were
taken on a freshly cleaved surfaces while data in (b)-(d) were
taken on aged (4 day old) surfaces. (d) EDCs at the $\Gamma$ point
taken with various photon energies} \label{fig1}
\end{figure*}

Single crystals were grown by a self flux technique, following the
previously reported recipe.\cite{Hor} ARPES measurements were
performed at the beam line 7U of UVSOR-II\cite{Ito}. Various
photon energies between 8 and 21 eV were used. The total energy
resolution was set to be 10 meV at 8 eV, and the angular
resolution was 0.1 degree. Samples were cleaved \emph{in situ} and
the chamber pressure was about 2$\times$10$^{-10}$ torr. The
measurement temperature was kept at 15 K.

%Fig1:
%Surface aging effect
ARPES data from Bi$_2$Se$_3$(111) are plotted in Figure 1. Panel
(a) shows data from freshly cleaved surface with 8 eV photon.
Relatively sharp and strongly dispersive electron pocket is
observed. This is the TM band having a helical spin
structure.\cite{Hsieh} As indicated by white arrow in the panel,
position of the TM band at the $\Gamma$ point (Dirac point) is
around 0.33 eV. One can also see broad weak feature near E$_F$ at
the $\Gamma$ point which originates from the bulk
states.\cite{Xia}

Panels (b) shows ARPES data taken with the same photon energy but
4 days after cleaving (aged surface). It is remarkable to note
that the TM band feature remains intact and looks almost the same
as that from a freshly cleaved surface. One difference is that the
Dirac point is now located at around 0.43 eV, about 0.1 eV higher
than the original position. This shift of the Dirac point is
attributed to charge transfer from physisorbed atoms or molecules.
While a similar surface doping effect in Bi$_{2}$Te$_{3}$ has
already been discussed,\cite{Chen,Noh,Hsieh3} no such effect has
been reported in Bi$_{2}$Se$_{3}$(111). The difference in the bulk
states other than the uniform shift is due to the $k_z$-selection
rule which will be discussed below.

%Determining bulk bottom binding energy
Figure 1c and 1d plot the data from aged surface taken with 10 and
17 eV. Compared to 8 eV data, they have almost the same TM
features while bulk bands look quite different. To see the photon
energy dependence more clearly, we plot in figure 1(d) energy
distribution curves (EDCs) at the $\Gamma$ point from fresh and
aged surfaces taken with various photon energies between 8 eV and
21 eV. The bulk state feature in the EDC taken with 8 eV photon is
relatively sharp. This feature becomes broader as photon energy
increases. This can be explained by the decreasing photoelectron
life time with increasing photon energy, resulting in inclusion of
a broader range of k$_z$.\cite{Smith} This, in combination with
the fact that probed k$_z$ varies with the photon energy, results
in clear observation of the bulk conduction band bottom (BCBB)
edge at about 0.2 eV in the EDCs taken with 17, 19 and 21 eV
photons as indicated by the dashed line in Figure 1(d) while 8 eV
EDC does not display a clear BCBB edge. In addition, 10 eV is the
right photon energy to observe the BCBB as can be seen in Figure
1(b) and 1(d). Considering the fact that there is 0.1 eV shift in
the band position between fresh and aged surfaces, we can
determine that BCBB of the fresh surface is at about 0.1 eV.

%surface electrons scattering channels with bulk electrons
We now wish to analyze the self energy of the TM states. Before we
proceed to analyzing the actual data, however, we need to touch
upon various scattering channels. In estimating the self energy,
we note that scattering between TM electrons should be small due
to the limited number of states (compared to the bulk states) and
helical spin structure.\cite{Roushan} Therefore, the dominant
mechanism for TM electron (or hole) decay is through transition to
bulk states. Figure 2(a) shows possible scattering channels
between a hole in the TM states and a bulk electron. A hole in the
TM state may decay into a bulk electronic state with the total
energy and momentum conserved through electron-hole (e-h) pair
creation, phonon emission or absorption, and impurity created
potential as illustrated in Figure 2(a). A rough estimate of the
imaginary part of the TM electron self-energy (or 1/$\tau$) can be
made by calculating the available phase space volume. We assume
that phonons between 0 and 0.03 eV equally contribute to
scattering and also that the bulk density of states is
proportional to $(E-E_{BB})^{1/2}$ where $E_{BB}$ is the BCBB
energy. A schematic of the result at 0 K is shown in Figure 2(b).
While Im$\Sigma$ in the Fermi liquid theory\cite{Claessen}
increases proportional to $\omega^2$, Im$\Sigma$ due to e-h pair
creation in Bi$_2$Se$_3$ increases at low binding energy side and
saturates slightly after E$_{BB}$. On the other hand, phonons and
impurities provide major scattering channels for the states with
binding energies lower than E$_{BB}$. These channels however do
not contribute for the states with binding energies higher than
E$_{BB}$.

\begin{figure}
\centering \epsfxsize=6.0cm \epsfbox{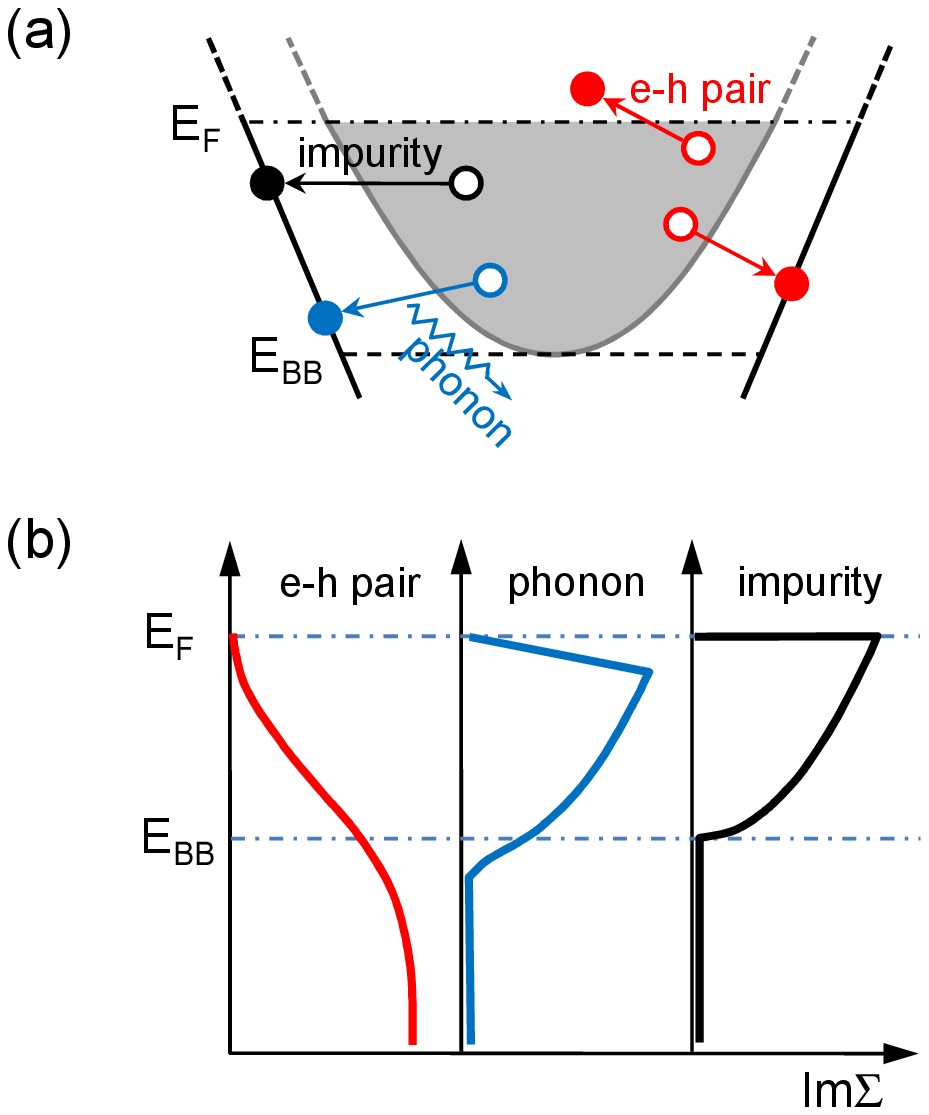} \caption{(a)
Schematic of various scattering channels for a photo-hole in the
TM band. Only the transition to the bulk states are considered.
(b) Corresponding Im$\Sigma$ for various channels.} \label{fig2}
\end{figure}

%How we extract self-energy from ARPES data
%Difficulty of self energy analysis on this system and advantage of
%8eV photon energy to do self energy analysis
ARPES spectral function $A(k,\omega)$ is proportional to the
imaginary part of the Greens function
\begin{eqnarray}
ImG(k,\omega)=\frac{Im\Sigma(k,\omega)}{(\omega-\varepsilon_{k}-Re\Sigma(k,\omega))^{2}+Im\Sigma(k,\omega)^{2}}.
\end{eqnarray}
If the self energy $\Sigma$(k,$\omega$) is slowly varying with
$\omega$, EDCs are Lorentzians and the half width at half maximum
(HWHM) of each EDC gives Im$\Sigma$. Extracted EDC peak width has
been used to study the electron dynamics in Mo(111) surfaces
states and cuprates.\cite{Valla,Chul,Park} Fitting an EDC with a
Lorentzian function however is often difficult in correlated
materials due to, for example, the incoherent spectral weight.
Instead, a commonly used method is to obtain Im$\Sigma$ by
multiplying HWHMs of momentum distribution curves (MDCs) by the
band velocity,\cite{Bogdanov,Zhou} which we also use in analyzing
the data. Another note is that it helps to isolate the TM
contribution of the spectral function. In the case of Bi$_2$Se$_3$
(111), bulk bands are located close to the TM band, making it
difficult to extract the TM spectral weight from ARPES data
precisely. However, we note that the bulk state spectral function
taken with 8 eV photon is much more suppressed than other data
sets as shown in Figure 1 (a)-(c). This is, as discussed above,
k$_z$-selection rule suppresses photoemission from the bulk states
at this photon energy. This provides us an opportunity to extract
the TM spectral function and do self energy analysis more
reliably. We thus analyze and present self energy analysis on 8 eV
data.

%Figure 3(a): Scattering rate start falling bellow E_{BB}
Figure 3(a) plots Im$\Sigma$ from fresh and aged surfaces.
Im$\Sigma$ initially increases but starts to decrease at around
0.10 eV and 0.17 eV binding energies for fresh and aged surfaces,
respectively. In addition, the fresh surface data hints flattening
of Im$\Sigma$ curve from 0.25 eV binding energy. These behaviors
contrast with ordinary metallic cases where Im$\Sigma$
monotonically increases due to electron-electron
scattering.\cite{Claessen,Valla,Park,Bogdanov,Zhou} The decrease
in Im$\Sigma$ reminds us of the drop in Im$\Sigma$ from phonon and
impurity channels at E$_{BB}$. Thus we can attribute the
Im$\Sigma$ behavior to the fact that TM electrons are dominantly
scattered to bulk electronic states which exist only below
E$_{BB}$.

\begin{figure}
\centering \epsfxsize=6.0cm \epsfbox{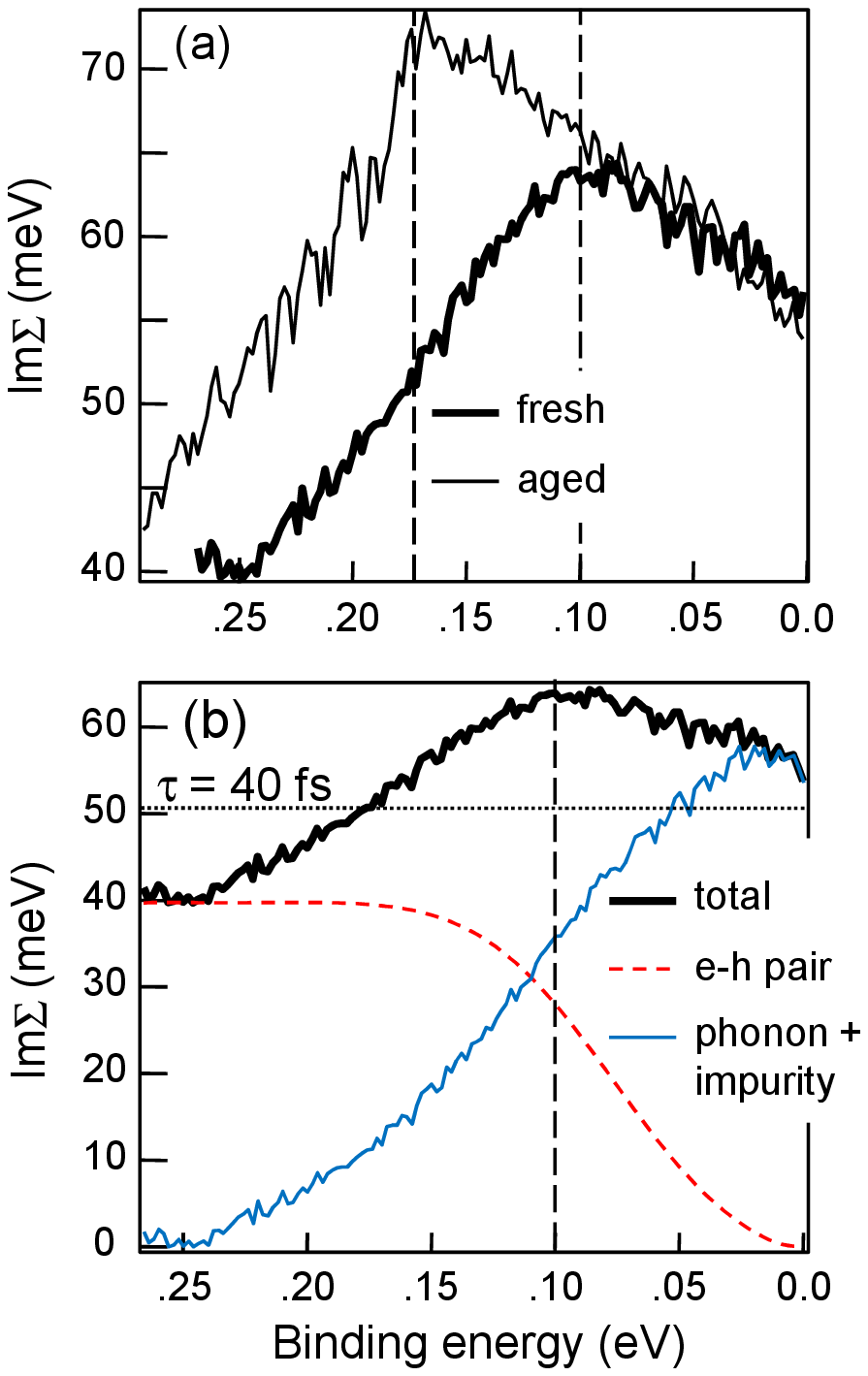} \caption{(a) TM
band Im$\Sigma$ from fresh and aged surfaces of
Bi$_{2}$Se$_{3}$(111) taken with 8 eV photons. (b) Analysis of
Im$\Sigma$ from fresh surface in terms of e-h pair creation, and
phonon plus impurity channels.} \label{fig3}
\end{figure}

%Figure 3(b): Impurity (or defect) will be main harm
We analyze the experimental Im$\Sigma$ of the 8 eV data from fresh
surfaces based on the interpretation illustrated in Figure 2. The
results are plotted in Figure 3(b). It is seen from the result
that, while Im$\Sigma$ at high energy side mainly come from
electron-electron interaction (electron-hole pair creation)
channel, impurity and phonon channels dominate the low energy
dynamics which controls the transport properties. As contributions
from both impurity and phonon channels are approximately
proportional to the bulk density of states, it is difficult to
distinguish them. However, we believe that the low energy part of
Im$\Sigma$ mainly comes from the defect scattering because
characteristic phonon kink in Im$\Sigma$ illustrated in Figure
2(b) was not observed. This is not unreasonable considering the
fact that these materials tend to be non-stoichiometric. This will
be the main obstacle to overcome if dissipationless spin current
on TIs were to be realized. The life time ($\tau$) and mean free
path $(l_m)$ of the electrons near $E_F$ are roughly $\tau$=40 fs
and $l_m=\tau\times\nu_g$=$0.02\mu$m, respectively.

As discussed above, decay of quasi-particles in TM states appear
to mainly come from scattering to the bulk states, not to other TM
states. Indeed, there are indications in other ARPES data that the
intrinsic life time of quasi-particles in TM bands is fairly long.
For example, Bi$_{2}$Te$_{3}$ shows a relatively sharp ARPES
lineshape.\cite{Chen}. Even though not an insulator, Sb$(111)$
also has TM states with very sharp ARPES peaks\cite{Gomes}. For
the low energy electron dynamics, we showed that defect scattering
plays the main role. It is therefore important to make
stoichiometric Bi$_{2}$Se$_{3}$ not only to make insulating bulk
states but also to reduce the scattering between TM and bulk
states.

Finally, we discuss if the TM states are protected. We observe
that the Fermi surface volume of the TM band from aged surfaces
increased about 2.3$\%$ compared to that of the fresh surfaces,
which means that at least 0.023 electrons per surface unit cell
are transferred presumably from adsorbed atoms or molecules on the
sample surface. Assuming that each atom or molecule donates 1
electron and that all the donated electrons are localized at the
surface states, there are $6\times 10^{13}/cm^2$ adsorbate
atoms/molecules which translates into $\approx 13 \AA$
inter-adsorbate distance. Considering the fact that bulk states
also receive electrons, the inter-adsorbate distance should be
shorter. Since the adsorbates should induce disorder potential
near the surface, additional scattering channels between surface
electrons in ordinary cases. Therefore, we expect an increase in
Im$\Sigma$ near E$_{F}$. However, we do not observe such increase
in the data in Figure 3(a). Please note that the mean free path
estimated from the life time in Figure 3(b) is about 200 $\AA$,
much longer than the estimated inter-adsorbate distance.

There can be different reasons why the scattering rate did not
increase. Interesting case would be from the protected nature of
the TM states. The estimated number of adsorbate atoms or
molecules on the surface may be enough to increase the scattering
rate substantially. However, the scattering rate may have not
increased because the TM states are topologically protected from
weak disorder from the potential induced by adsorbates. If it is
indeed the case, this would be the first experimental evidence
from ARPES for the protected nature of TM states, to our best
knowledge. However, one has to be careful not to make a foregone
conclusion because the induced potential can vary depending on the
adsorbates. For example, relatively small amount of H atoms on
graphene destroys the metallic states while K atoms have almost no
effect except electron doping.\cite{Bostwick} In order to resolve
this issue, we propose a controlled ARPES experiments in
combination with first principles electronic structure
calculations.

Authors acknowledge fruitful discussions with J.H.Han. This work
is supported by NRF (20090080739) and by KOSEF
(F01-2007-000-10117-0). SRP is supported by the 2009 Yonsei
University Research Fund. NH acknowledges support from Korea
Research Foundation through KRF-2008-331-C00093

\end{document}